\documentclass[onecolumn]{aastex63}
\usepackage{xspace}
\usepackage{multirow}
\received{December 23, 2020}
\revised{January 21, 2021}
\accepted{\today}
\submitjournal{ApJL}

\shorttitle{Bidirectional magnetoacoustic wave}
\shortauthors{Miao et al.}

\newcommand{\tabref}[1]{Table~\ref{#1}}
\newcommand{\secref}[1]{Section \ref{#1}}
\newcommand{\figref}[1]{Figure~\ref{#1}}
\newcommand{\kms}{\ensuremath{\,\mathrm{km}\cdot \mathrm{s}^{-1}}}
\newcommand{\unit}[1]{\ensuremath{\,\mathrm{#1}}}
\newcommand{\degree}{\ensuremath{^\circ}}
\newcommand{\Alfven}{Alfv\'en\xspace}
\newcommand{\Alfvenic}{Alfv\'enic\xspace}
\usepackage{lineno}
\begin{document}

\title{Diagnosing a Solar Flaring Core with Bidirectional Quasi-Periodic Fast Propagating Magnetoacoustic Waves}

\author[0000-0003-2183-2095]{Yuhu Miao}
\email{miaoyuhu@hit.edu.cn}
\affiliation{Institute of Space Science and Applied Technology, Harbin Institute of Technology, Shenzhen, Guangdong 518055, China}
\affiliation{CAS Key Laboratory of Solar Activity, National Astronomical Observatories, Beijing 100012, China}

\author{Dong Li}
\affiliation{Key Laboratory of Dark Matter and Space Astronomy, Purple Mountain Observatory, CAS, Nanjing 210033, China}

\author[0000-0002-9514-6402]{Ding Yuan}
 \email{yuanding@hit.edu.cn}
\affiliation{Institute of Space Science and Applied Technology, Harbin Institute of Technology, Shenzhen, Guangdong 518055, China}

\author{Chaowei Jiang}
\affiliation{Institute of Space Science and Applied Technology, Harbin Institute of Technology, Shenzhen, Guangdong 518055, China}

\author[0000-0002-5391-4709]{Abouazza Elmhamdi}
\affiliation{Department of Physics and Astronomy, King Saud University,
PO Box 2455, Riyadh 11451, Saudi Arabia}

\author{Mingyu Zhao}
\affiliation{Yunnan Observatories, Chinese Academy of Sciences, Kunming 650011, China}

\author{Sergey Anfinogentov}
\affiliation{Institute of Solar-Terrestrial Physics, 664033, Irkutsk, Russia}

\begin{abstract}

Quasi-periodic fast propagating (QFP) waves are often excited by solar flares, and could be trapped in the coronal structure with low \Alfven speed, so they could be used as a diagnosing tool for both the flaring core and magnetic waveguide. As the periodicity of a QFP wave could originate from a periodic source or be dispersively waveguided, it is a key parameter for diagnosing the flaring core and waveguide. In this paper, we study two QFP waves excited by a {\em GOES}-class C1.3 solar flare occurring at active region NOAA 12734  on 8 March 2019. Two QFP waves were guided by two oppositely oriented coronal funnel. The periods of two QFP waves were identical and were roughly equal to the period of the oscillatory signal in the X-ray and 17 GHz radio emission released by the flaring core. It is very likely that the two QFP waves could be periodically excited by the flaring core. Many features of this QFP wave event is consistent with the magnetic tuning fork model. We also investigated the seismological application with QFP waves, and found that the magnetic field inferred with magnetohydrodynamic seismology was consistent with that obtained in magnetic extrapolation model. Our study suggest that the QFP wave is a good tool for diagnosing both the flaring core and the magnetic waveguide.
\end{abstract}

\section{Introduction}

Quasi-periodic fast propagating magnetoacoustic (QFP) wave was first reported by \citet{liuw11} with full-disk imaging capability of the Atmospheric Imaging Assembly \citep[AIA;][]{lemen12} onboard the \textit{Solar Dynamics Observatory} \citep[\textit{SDO};][]{pesnell12}. QFP wave propagates at local \Alfven speed across a distance at the scale of solar radius. Its periodicity appears to be consistent with pulsating period of the light emission flux of solar flare, therefore, it is believed to be excited by repetitive flaring energy releases \citep{liuw12,shen12b,shen2013b}.

\citet{yuanding2013} found that the QFP wave trains were likely to be triggered by spiky flaring energy releases, and they suggested that QFP wave trains could be excited impulsively and evolve into a quasi-periodic nature with the wave guiding effect of magnetized coronal structure. This scenario was demonstrated by injecting impulsive energy into a diverging magnetic funnel. The quasi-periodic nature was well reproduced both in the trapped fast magnetoacoustic wave and the leaky mode \citep{pascoe2013, pascoe2014b,quzhining2017}. A number of studies supported the dispersive evolution of fast magnetoacoustic wave within coronal waveguides \citep{nistico2014,shenyd2018,miao2019,miao2020}.

\citet{goddard2019} did a parametric study on the initial impulsive drive of QFP wave and found that the final spatial and spectral signatures of the guided QFP wave trains depend strongly on the temporal duration of the initial perturbation. This dependence gives rise to the potential of diagnosing the flare core region with the exited QFP wave. \citet{takasao2016} did a numerical magnetic reconnection experiment and showed that the flaring loop is bombarded by the backward flow from reconnecting site. Such a recurring plasma flow created a magnetic tuning fork above the loop top. This region acts as an \Alfvenic resonator and becomes a source of quasi-periodic process, therein the QFP wave could leak from the loop top. This scenario was also justified with SDO/AIA's multi-temperature observations \citep{takasao2016,lileping2018}

QFP wave have potential application in diagnosing the flaring core and its magnetic waveguide. \citet{goddard2016} found that QFP wave trains could overrun the leading edge of coronal mass ejection (CME), their interaction could modulate the radio emissions and generate quasi-periodic sparks in the radio spectrograph. This signal could supply the key information of CME's expanding front. \citet{ofman2018} provide the first evidence of counter-streaming QFP waves that could potentially lead to turbulent cascade and dissipate sufficient energy flux for coronal heating in low-corona magnetic structures.

In this study, we aim to constrain the source of periodicity with bidirectional QFP wave excited by a flare. Both the wave train and the flaring energy release exhibited identical periodicity. We propose to employ the potential of bidirectional QFP wave to constrain the origin of periodicity, and to investigate the capability of plasma diagnostics. In \secref{sec:instr}, we describe the flare-triggered QFP event and the relevant data analysis. \secref{sec:obs} presents the main results; the discussion and conclusions are given in \secref{sec:sum}.

\section{Observations and Data analysis}
\label{sec:instr}

\begin{figure*}
	\epsscale{1.0} \plotone{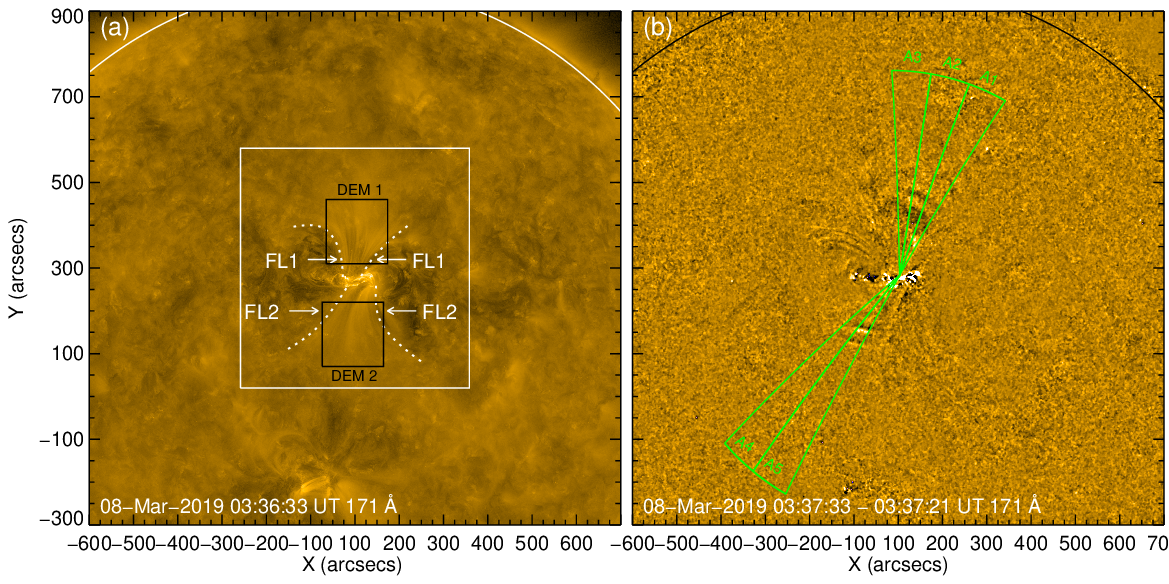}
	\caption{(a) AIA 171 \AA{} image showing AR 12734 and the coronal funnels. Regions enclosed by DEM1 and DEM2 were used in the DEM analysis. FL1 and FL2 label two coronal funnels used in this study. The white dotted lines indicate the borders of FL1 and FL2. The white rectangle indicate the region used in DEM analysis as shown in \figref{fig:fig5}. (b) Difference image of the AIA 171 \AA{} channel to highlight wave propagation. Sectors A1-A5 were used to make time-distance plots shown in \figref{fig:fig2}{} (refer to
animation.mpeg). \label{fig:fig1}}
\end{figure*}

\begin{figure*}
	\epsscale{1.0} \plotone{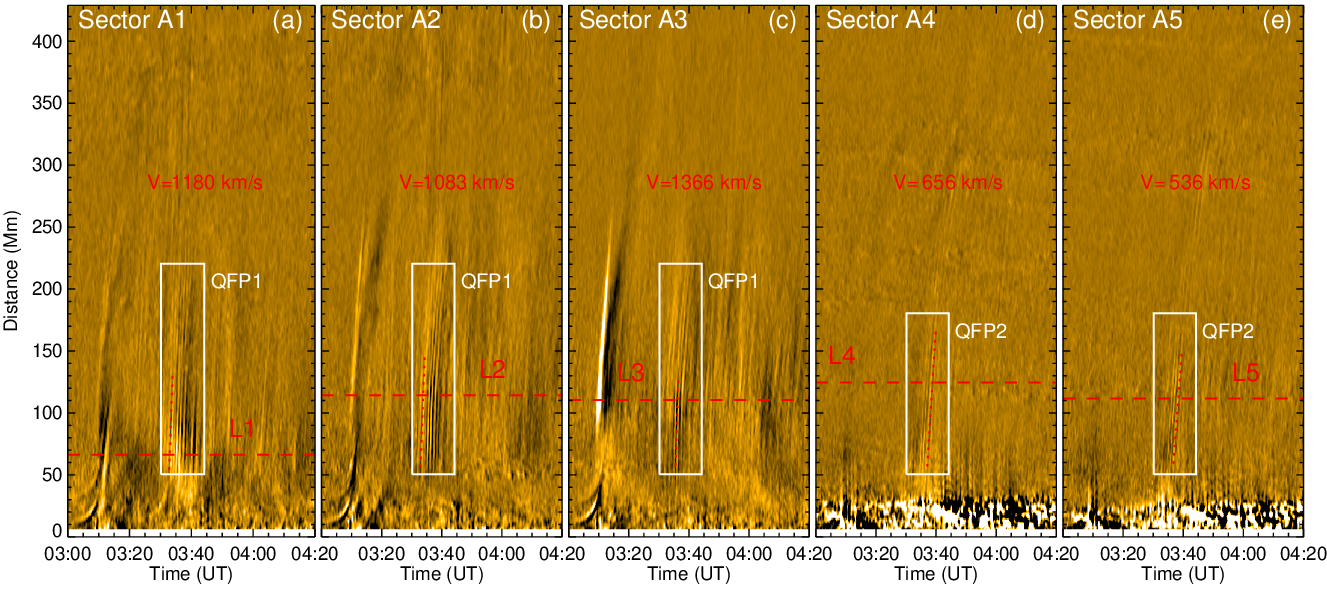}
	\caption{ (a)-(e) time-distance diagrams corresponding to sectors A1-A5. The QFP wave features are indicated by the boxes.  The red-dashed lines mark the positions used in wavelet analysis for period measurement. Each red dotted line within a white rectangle follows a sample QFP wave front, its slope was used to measure the propagating speed.\label{fig:fig2}}
\end{figure*}

\begin{figure*}
	\epsscale{0.9} \plotone{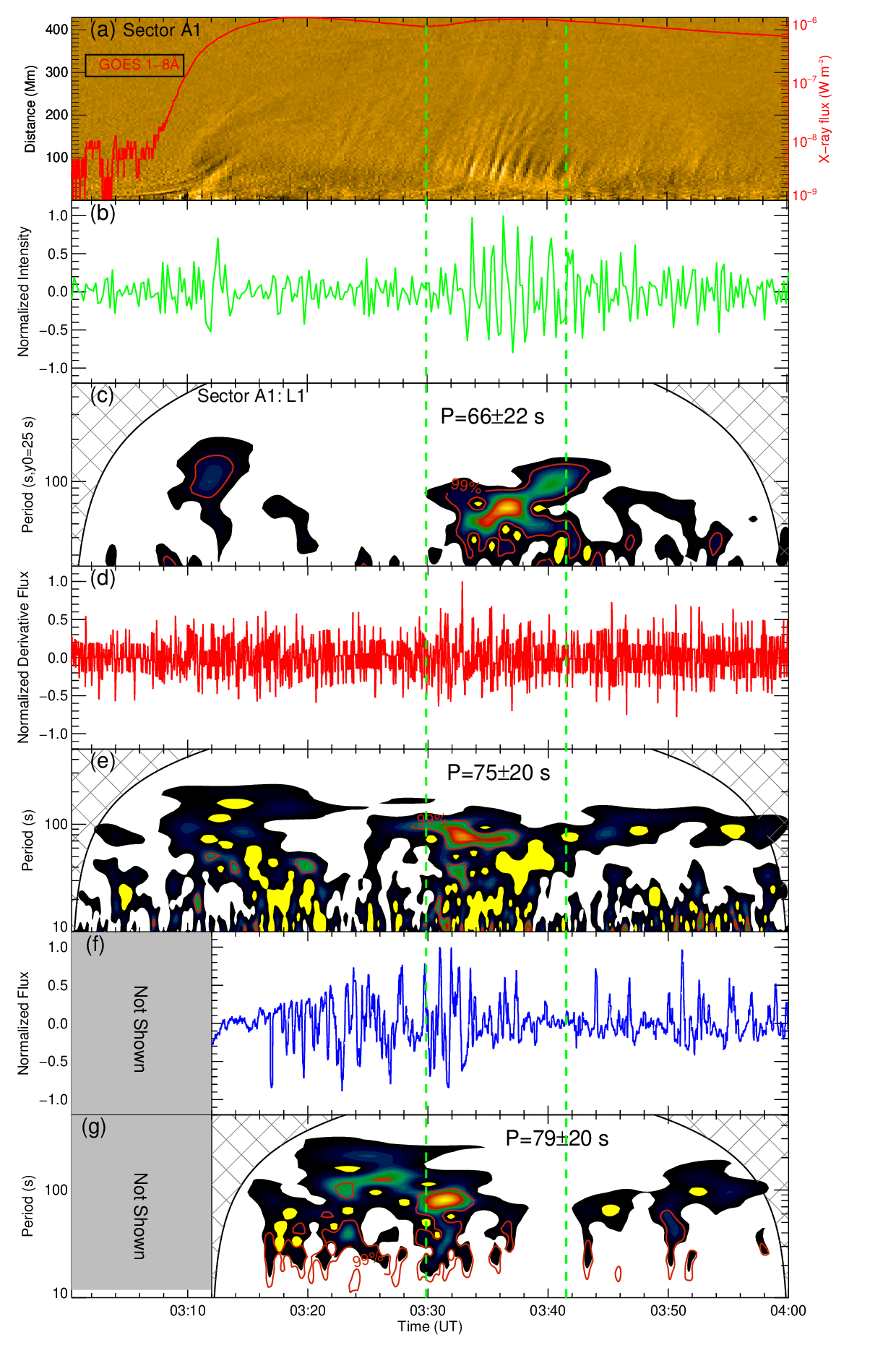}
	\caption{(a) Time-distance plot of Sector A1, overlaid with the GOES X-ray flux. (b) and (c) Detrended emission intensity measured at L1 in \figref{fig:fig2}(a) and its wavelet spectrum. (d) and (e) Derivative of GOES X-ray flux and its wavelet spectrum. (f) and (g) Detrended RoRH 17 GHz radio emission flux and its wavelet spectrum. The radio data of NoRH 17 GHz started at about 03:12 UT. The two green vertical dashed lines highlight the time interval of QFP wave and periodic signal in the flaring core. The y0 indicates the initial value of the Y-axis in panel (c). \label{fig:fig3}}
\end{figure*}

\begin{figure*}
	\epsscale{1.0} \plotone{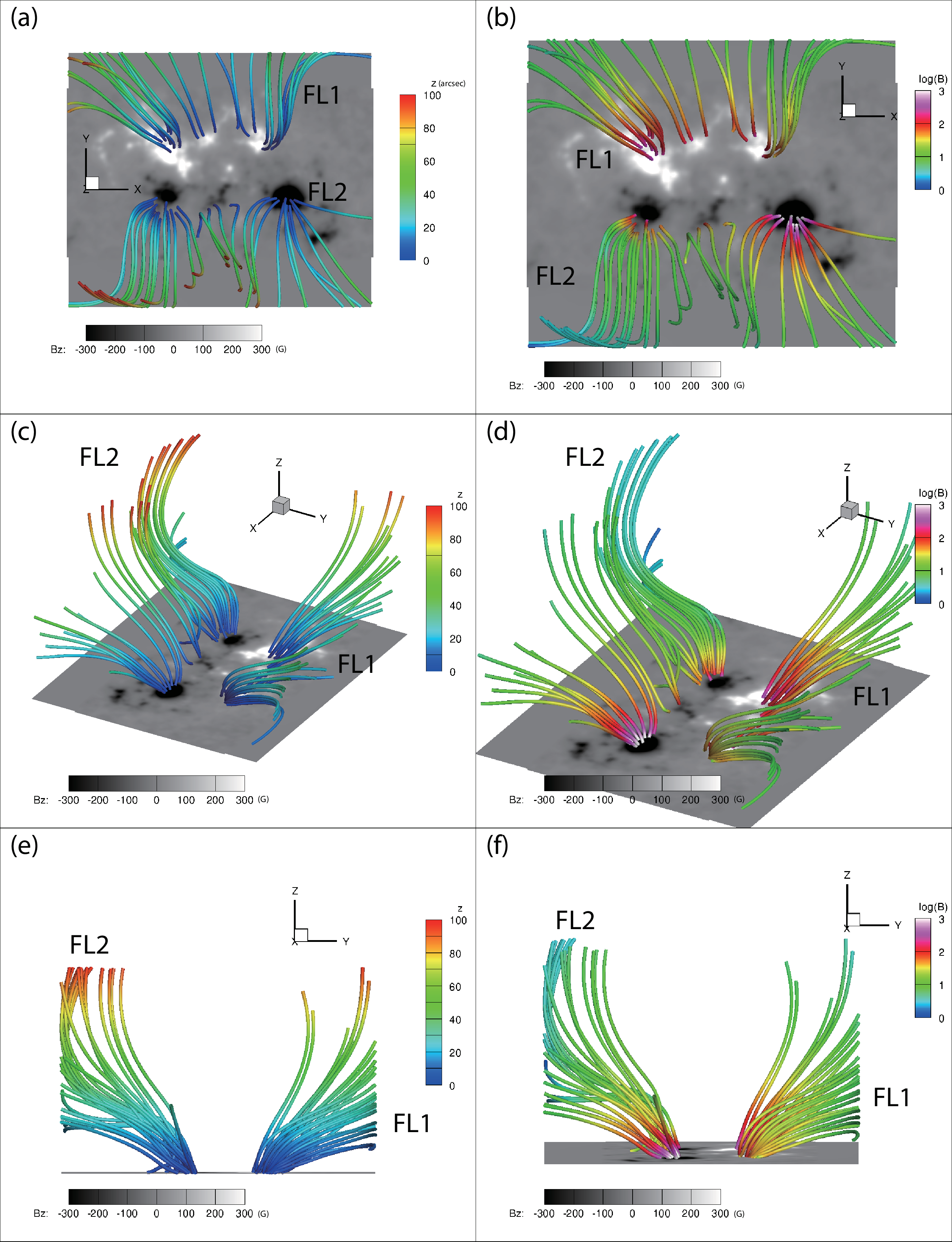}

	\caption{Non-linear force-free magnetic field extrapolation. (a)-(f) shows magnetic field line highlighting coronal funnels FL1 and FL2 at different viewing angle. Left column uses height as color bar to show the magnetic field lines, whereas right column uses field strength to label color. The FOV of the magnetogram is about 340\arcsec $\times$ 200\arcsec. \label{fig:fig4}}
\end{figure*}

\begin{figure*}
	\epsscale{1.0} \plotone{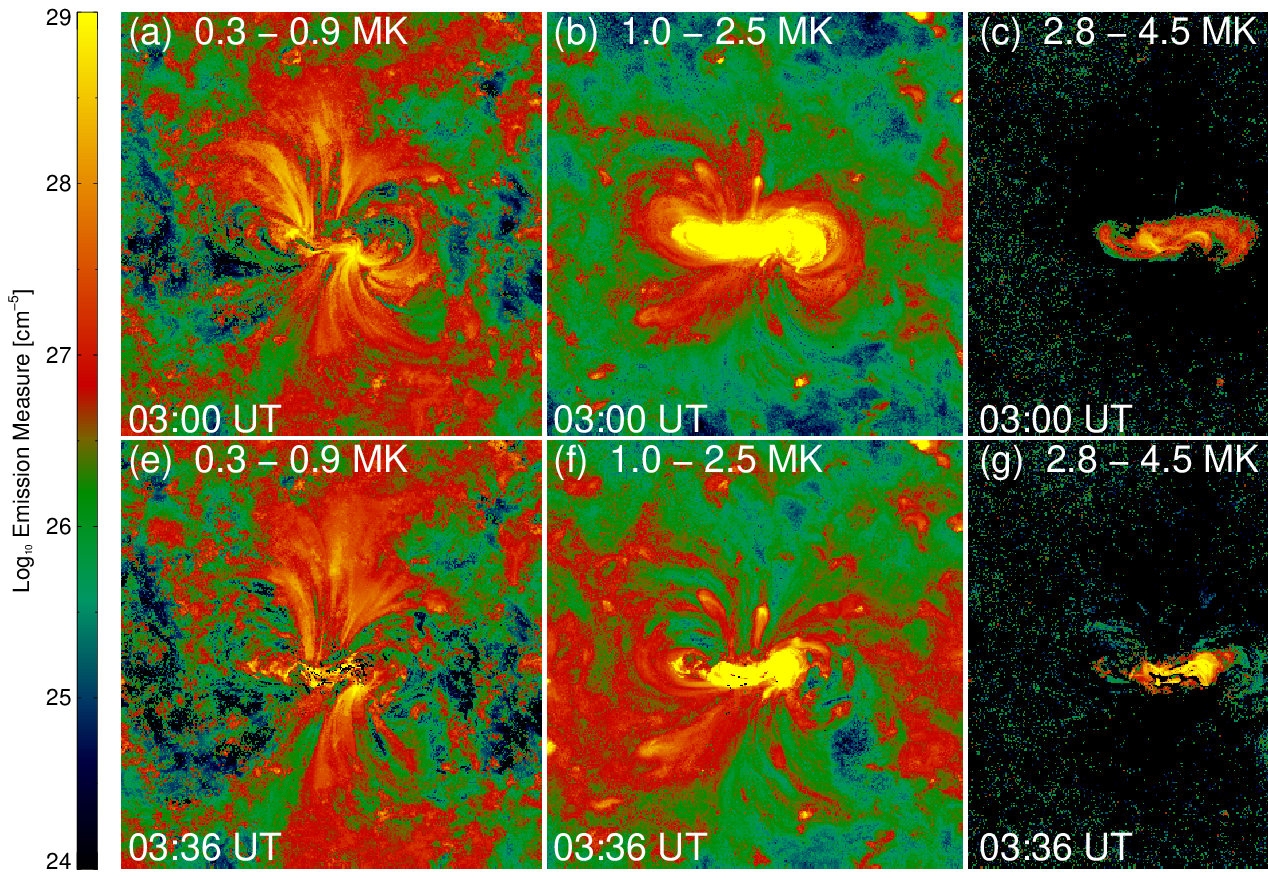}
	\caption{DEM maps as a function of temperatures.  (a)-(d)  show the DEM maps before the flare eruption at 03:00 UT. (e)-(h) present DEM maps during the QFP wave propagation at 03:36 UT.
		\label{fig:fig5}}
\end{figure*}

In this study, we analysed a C1.3-class solar flare and the associated wave features originating from active region NOAA 12734 on 08 March 2019. The integrated X-ray flux over the range of 1 \AA{} - 8 \AA{} was recorded by {\em Geostationary Operational Environmental Satellite} ({\em GOES}). The {\em GOES} X-ray flux started to rise at 03:07 UT, reached maximum at 03:18 UT, and dropped off at around 04:00 UT. This flare had two emission peaks. During the first flaring peak, the eruption of a higher-lying filamentary structure excited a large-scale extreme ultra-violet (EUV) wave \citep[see reviews by ][]{chenpf2011,liuw2014,warmuth2015}. This EUV wave propagated radially from the flare epi-center, with the projected phase speed at about $200\kms$-$600\kms$. The second flaring peak triggered a pair of QFP waves. These two QFP waves propagated outwardly following two coronal funnels oriented to the north and south, respectively. The flaring process and the associated wave excitation is illustrated in \figref{fig:fig1} (see animation.mpeg).

The C1.3-class solar flare and the associated EUV and QFP waves were recorded by SDO/AIA. We used the AIA 171 \AA{} and 193\AA{} data to study this event. The AIA data were calibrated with the standard routine provided by the Solar Software (SSW), each image was normalized by its exposure time. An AIA image pixel corresponds to an angular width of $0\arcsec.6$; the image sequences in all AIA channels had a cadence of about 12 seconds.

To quantify the kinematics of the bidirectional QFP waves, we took the intensities along a slit with the origin located at the flare epi-center. To improve the signal-to-noise ratio, we averaged the intensity with a few neighbouring pixels, the number of pixels used for the averaging was proportional to the distance to the flare epi-center. This technique is indicated by a set of five sectors labelled with ``A1-A5'' in \figref{fig:fig1}, each sector had an angular extent of 10 degrees. The sectors were chosen to sample the QFP waves with significant amplitude. The spatially averaged intensities was stacked in order of time to form time-distance plots as shown in \figref{fig:fig2}. To highlight the wave propagation feature, we used the running-difference images. Each difference image was calculated by taking the backward difference with the image taken 12 seconds earlier.

To reveal the oscillatory signal of the QFP wave, we performed wavelet analysis \citep{torrence1998,wangzhengkai2020,liangbo2020,fengsong2020b} to the intensity variation at the position L1-L5 at sector A1-A5 (see \figref{fig:fig2}). We used the same method to analyse the period of the QFP wave train observed in the AIA 193 \AA{} channel, (see \tabref{tab:QFP}). We also analysed the derivative of GOES X-ray flux as well as the Nobeyama Radioheliograph (NoRH) 17 GHz data. The wavelet spectra are illustrated in \figref{fig:fig3}.

The QFP waves appeared to be guided by the open magnetic funnel, therefore, we did a nonlinear force-free field (NLFFF) extrapolation to show the magnetic configuration of AR 12734 \citep{jiangchaowei2018,zoupeng2020}. The boundary magnetogram  was provided by SDO/HMI. We visualized the magnetic field lines from strong polarities at the active region 12734 to highlight funnel structures as displayed in \figref{fig:fig4}.

To estimate the plasma temperature and density of the coronal funnel, we used six EUV channels of SDO/AIA to calculate the differential emission measure (DEM), the code was developed by \citet{cheung2015} and \citet{suyang2018}. The DEM inversion was done at 03:00 UT before the QFP wave excitation and 03:36 UT during this process. The DEM maps at different temperature ranges are plotted in \figref{fig:fig5}, which reveals the plasma emissions before and during the QFP wave propagation. To evaluate the plasma temperature and density of the coronal funnel, we selected two regions, labelled as DEM1 and DEM2 in \figref{fig:fig1} {(a)} and integrated the average DEM to calculate the emission measure (EM). The thickness of a coronal funnel was estimated within the magnetic field extrapolation model, it was about three times of a typical width of a coronal loop observed at AR 12734. This thickness was used as an estimate of the column depth ($d$) of the coronal plasma. Then, the electron density could be estimated using $n=\sqrt{\mathrm{EM}/{d}}$.

\section{Results}
\label{sec:obs}
\subsection{Kinetics of the bidirectional QFP waves}

\figref{fig:fig2} reveals the propagation and periodicity of the bidirectional QFP waves. Section A1-A3 was placed over coronal funnel 1 (FL1) and have captured the feature of QFP wave 1 (QFP1). We could see that the QFP1 wave propagated at the speeds of $1083\,\kms$  to $1366\kms$. The periods obtained at position L1, L2, and L3 were $66\pm22\unit{s}$, $63\pm21\unit{s}$, and $62\pm22\unit{s}$, respectively. These periodicities were measured in the wavelet spectrum, as shown as an example in \figref{fig:fig3}.

Sectors A4 and A5 captured the features of QFP wave 2 (QFP2) propagating in coronal funnel 2 (FL2) in the AIA 171 \AA\ channel. Their speeds were measured to be $656\kms$ and $536\kms$ at sectors A4 and A5, respectively. The periodicities at these two sectors were $65\pm21\unit{s}$, $66\pm19\unit{s}$, respectively. In the 193 \AA\ channel, the projected propagation speeds were relatively smaller, the respective speeds were about $397\kms$ and $358\kms$; the corresponding periods were $73\pm19\unit{s}$ and $66\pm20\unit{s}$, respectively. The propagation speeds and periods are listed in Table~\ref{tab:QFP}.

\subsection{Periodicities of flare emission and QFP waves}

The bidirectional QFP waves followed two separate coronal funnel oriented towards opposite directions. It appears that the two QFP waves exhibited identical periodicities, which could have a common origin from the flaring core. We use the light curve of soft X-ray emission and the 17 GHz radio emission integrated over AR 12734 to show the periodicity of the flaring core. In order to suppress the low-frequency spectral components,
we removed the general trend in the 17 GHz radio emission signal, whereas for the X-ray emission flux, we take its time derivative for further analysis. Therefore, the periodicities at minute timescale could be easily detectable. 

\figref{fig:fig3} implies that the detrended 17 GHz radio emission signal and the time derivative of the X-ray emission flux exhibited a periodicity at about one minute, the periods were about $75\pm20\unit{s}$ and $79\pm20\unit{s}$, respectively. The periodic signal in the 17 GHz radio flux started to oscillate at about 03:29 UT and disappeared at about 03:34,  the duration of periodicity was about 5 minutes, see \figref{fig:fig3}(g). The oscillatory signal in the GOES X-ray emission started at about 03:32 UT, about 3 minute lagged behind the radio signal, see \figref{fig:fig3}(d). We shall note here that the GOES X-ray flux and the periodic fast wave started almost simultaneously.

\subsection{Magnetic structure of the coronal funnels}
\label{sec:mag}

\figref{fig:fig4} displays the magnetic field lines originating from the strong polarities at AR 12734, where the magnetic structures of two coronal funnel were clearly seen. FL1 was about 30\degree deviated from the plane of the boundary, whereas FL2 have a deviation angle of about 45\degree. We shall note that AR 12734 has an altitude of about 20\degree. Henceforth, after correcting the projection effect, we obtained that FL1 and FL2 deviated from the plane-of-sky for about $\phi_1=10\degree$ and $\phi_2=65\degree$, respectively. The magnetic field strength was estimated at $B_1^{\mathrm{extr}}=(7-23)\unit{G}$ for FL1 and  $B_2^{\mathrm{extr}}=(4-22)\unit{G}$ for FL2.

Here we used extrapolated magnetic field to estimate the width of  two coronal funnels, and found that both of them had a width of about 5 Mm, this value is about two to three times of the typical width of a coronal loop in this active region. This value could be used as an estimate of the column depth for the coronal funnels, however, we shall consider the line-of-sight effect. Therefore, the column depth for FL1 is about $d_1=5.1\unit{Mm}$, and $d_2=11.8\unit{Mm}$ for FL2.

\subsection{Plasma temperature and density}

\figref{fig:fig5} shows DEM maps at various temperature ranges from 0.3 MK to  8 MK. The flaring core was consist of  hot plasma at a broad range of temperature from 1 MK to 8 MK. We shall note that during the QFP wave process, flaring core was heated to high temperature as a whole (see \figref{fig:fig5}{(h)}), this pattern had big contrast to the status before the QFP process, during which the flaring core had only hot plasma in the filamentary structure, see \figref{fig:fig5}{d}.  Two coronal funnel confined plasma mostly at temperatures at 0.3 MK to 0.9 MK.

The plasma temperature was calculated by locating the temperature where DEM reached its maximal value. FL1 and FL2 had a average temperature of about $T_1=1.23\unit{MK}$ and $T_2=1.15\unit{MK}$, respectively.

We then integrated the DEM over temperature and obtained the emission measure (EM). The DEM had two district peaks at low and high temperature ranges, the high temperature component may arise owing to the contribution of low-lying plasma structure, such as those see in \figref{fig:fig5} (b) and (f). So, when we calculated the EM, the high temperature component was disregard as it was done in \citet{lidong2020a}. We find that
FL1 had an average electron number density at about $4.86\times10^8\unit{cm^{-3}}$ before the QFP and dropped to $4.18\times10^8\unit{cm^{-3}}$ when QFP wave started to propagate along it. With the identical method, FL2 had a plasma with the average number density of electrons at about $2.65\times10^8\unit{cm^{-3}}$ before the flare and decreased to $2.45\times10^8\unit{cm^{-3}}$.

\subsection{Seismological application}
\label{sec:alfen}

QFP wave propagated at an average speed of about $v_1=1210\kms$ at FL1, if we correct the projection effect, the fast magnetoacoustic speed at FL1 was about $v_\mathrm{F1}=v_1/\cos\phi_1=1230\kms$. As FL1 had a plasma temperature of $T_1=1.23\unit{MK}$, the acoustic speed was $C_1=163\kms$ at coronal funnel FL1. Henceforth,  the \Alfven speed at FL1 could be obtained, $V_{\mathrm{A}1}=1220\kms$. Here we assumed that the wave vector is parallel to the magnetic field vector. Considering $v_2=596\kms$, $\phi_2=65\degree$ and $T_2=1.15\unit{MK}$, we obtained that the acoustic speed and \Alfven speed at coronal FL2, are about $C_2=158\kms$ and $V_{\mathrm{A}2}=1410\kms$, respectively. In combination with the density estimates, the average magnetic field for FL1 and FL2 are about $B_1^{\mathrm{seism}}=12.8\unit{G}$ and $B_1^{\mathrm{seism}}=11.3\unit{G} $, respectively.  The detailed parameters of plasma and magnetic field are listed in \tabref{tab:DEM}.

\begin{table*}[ht]
	\centering
	\caption{Parameters of the bidirectional QFP waves}
	\label{tab:QFP}
	\begin{tabular}{ccccccc}
		\hline
		\hline
	
	Region &	Wave & Channel & Slit  &  Phase speed(\kms)&  Position &  Period (s) \\
		\hline
		\multirow{3}{*}{Coronal Funnel 1}
	 &	QFP1 & 171 \AA & A1 & 1180 & L1 & 66 $\pm$ 22 \\
	&	QFP1 & 171 \AA & A2 & 1083 & L2 & 63 $\pm$ 20 \\
	&	QFP1 & 171 \AA & A3 & 1366 & L3 & 62 $\pm$ 22 \\
	\hline
		\multirow{3}{*}{Coronal Funnel 2}
	 &	QFP2 & 171 \AA & A4 & 656  & L4 & 65 $\pm$ 21 \\
	&	QFP2 & 171 \AA & A5 & 536  & L5 & 66 $\pm$ 19 \\
	&	QFP2 & 193 \AA & A4 & 397  & \nodata & 73 $\pm$ 19  \\
	&	QFP2 & 193 \AA & A5 & 358  & \nodata & 66 $\pm$ 20  \\
	\hline
		\multirow{2}{*}{Flaring core}
	&	NoRH &    17 GHz & \nodata & \nodata  & \nodata & 79 $\pm$ 20  \\ 
	&	GOES & $(1 - 8)\unit{\AA{}}$ & \nodata & \nodata  & \nodata & 75 $\pm$ 20  \\ 
	
		\hline
	\end{tabular}
\end{table*}

\begin{table*}[ht]
	\centering
	\caption{Parameters of the plasma and magnetic field}
	\label{tab:DEM}
	\begin{tabular}{lc}
		\hline
		\hline
		 Parameter & Value  \\
		\hline

		 		\multicolumn{2}{c}{Coronal Funnel 1} \\
		 		\hline
	        	 Temperature [$\unit{MK}$]   &  $1.23$  \\
		 	      Number density of electrons   [$\unit{cm^{-3}}$] & $4.18\cdot10^{8}$\\
		 	      Wave speed [$\kms$] &  1210 \\
		 	      Fast speed [$\kms$]&  1230 \\
		 	     Acoustic speed [$\kms$] &  163\\
		 	    \Alfven speed  [$\kms$]& 1220\\
		 	 Angle with the boundary plane in extrapolation  & $30\degree$\\
		 	 Angle with the plane of sky  & $10\degree$\\
		 	 Magnetic field strength  (extrapolation)  [$\unit{G}$] & $7-23$ \\
		 	 Magnetic field strength (seismology)  [$\unit{G}$]  &  12.8 \\
		 	 \hline
		 	 	\multicolumn{2}{c}{Coronal Funnel 2} \\
		 	 \hline
		 	    Temperature  [$\unit{MK}$] &  $1.15$  \\
	 				Number density of electrons   [$\unit{cm^{-3}}$] & $2.45\cdot10^{8}$\\
	 				Wave speed [$\kms$] & 596  \\
	 				Fast speed  [$\kms$]&  1410\\
			     	Acoustic speed [$\kms$] &  158\\
				\Alfven speed  [$\kms$]&  1400\\
				Angle with the boundary plane in extrapolation  & $45\degree$\\
				Angle with the plane of sky  & $65\degree$\\
			Magnetic field strength  (extrapolation) [$\unit{G}$] & $4-22$ \\
			Magnetic field strength (seismology)  [$\unit{G}$]  &  11.3 \\

 		\hline

		\hline
	\end{tabular}
\end{table*}

\section{Discussion and Conclusions}
\label{sec:sum}

In this paper, we analyse the oscillatory processes accompanying a C1.3 solar flare. The flare light curves in X-ray and radio emission both exhibited an oscillatory signal. Two QFP waves were launched from the flaring core, and propagated along two oppositely oriented coronal funnel structures. We took advantage of the unique magnetic structuring and studied the origin of the periodic signals, and investigated the potential seismological applications on the magnetized plasma and the flaring source.

In this event, the flaring core was an energy source of the QFP wave trains, whereas the coronal funnels acted as waveguides for the fast MHD waves. Fast magnetoacoustic waves could propagate across magnetic field lines, and the wave energy could be trapped by regions with low \Alfven speed \citep[see][for example]{pascoe2013,pascoe2014b}. The QFP waves at two separate coronal funnels were found to have identical periodicity, see \tabref{tab:QFP}. The oscillatory signal of the flaring core was revealed in the X-ray and 17 GHz radio emission, with periods slightly greater than the period of the QFP waves. However, if we consider the uncertainties in the measurement, they could be considered as roughly equal. We should also bear in mind that the flare core generated spiky energy releases, which could make the spectral analysis less accurate. It should be noted that the difference in speeds between 171 and 193 \AA\ channels are listed in \tabref{tab:QFP}. This difference might be caused by the inclination angle of coronal funnels with different temperatures.

We explored the seismological potential with DEM analysis, magnetic field extrapolation and MHD wave analysis. DEM analysis provides the thermal parameters of the plasma confined by the coronal funnels, i.e., plasma density and temperature. With the wave parameters, we could infer the \Alfven speed and henceforth the magnetic field strength. With magnetic field extrapolation, we obtained a reference that could be used to assess the accuracy and robustness of MHD seismology with QFP wave. With the geometric and magnetic parameters of the coronal funnels measured in the extrapolation, we find that the magnetic field strength obtained with MHD seismlogy agreed with the value measured in the magnetic field extrapolation model within a factor of about 0.5 to 2. This result is consistent with seismology applications with kink wave \citep{2011ApJ...736..102A} and the numeric kink wave model \citep{2009ApJ...699L..72D}. We shall note that the magnetic field diverges with height, so the magnetic field strength obtained in MHD seismology is in fact the value averaged over the magnetic waveguide, therefore, the MHD seismology with QFP wave gives reasonable values.

Based on the above analyses, we could conclude that the QFP wave might be generated by an oscillatory signal at the flaring core. This scenario agrees in many aspects with the magnetic tuning fork model simulated by \citet{takasao2016}. The magnetic field at the coronal loop top could be bombarded by repetitive flow generated by magnetic reconnection above the loop arcade systems. A magnetic tuning fork is formed above the loop top, and become an \Alfvenic resonator. The trapped fluid energy bounces back and forth within the magnetic tuning fork, becoming a quasi-periodic signal. This is the source of periodicity at the X-ray and radio signals. A magnetic tuning fork could trap fast magnetoacoustic waves, and they could leak through the boundaries and become QFP waves guided by the coronal funnel that were rooted close to the magnetic tuning fork. We shall also note that the QFP waves could be generated directly by repetitive magnetic reconnections at the flaring core as observed by \citet{lileping2018}.

In this study, the bidirectional QFP waves provides a new possibility to diagnose the features of flaring core and coronal magntic structures. Combining with DEM inversion and magnetic field extrapolation, the bidirectional QFP waves could be used to probe the features of the plasma waveguides in the solar atmosphere. As imaging observations are difficult to  reveal the magnetic activities at the flaring core, MHD seismology  with the periodic signal of QFP wave could be an alternative method to study the magnetic reconnection process above the loop top. This method is transferable to stellar flare investigation.

\acknowledgments
The authors thank the referee for his/her valuable suggestions that improved the quality of the letter. We thank for the excellent data provided by the {\em SDO}, {\em GOES} and the Nobeyma.
Y.H.M. and D.Y. is supported by the National Natural Science Foundation of China (NSFC, 11803005, 41731067), the Shenzhen Technology Project (JCYJ20180306172239618), the Shenzhen Science and Technology Program (Group No.KQTD20180410161218820), the Chinese Postdoctoral Science Foundation (2020M681085) and the Open Research Program of the Key Laboratory of Solar Activity of Chinese Academy of Sciences (KLSA202110). D. L. is supported by the NSFC 11973092. C.W.J. is supported by the NSFC 41822404 and the Fundamental Research Funds for the Central Universities (Grant No. HIT.BRETIV.201901). A.E. extend his appreciation to the Deanship of Scientific Research at King Saud University for funding this work through research group No. (RG-1440-092). M.Y.Z. is supported by the NSFC 11973086. We also thank C. Torrence and G. Compo for providing the wavelet software that is available at \url{http://atoc.colorado.edu/research/wavelets}.


\end{document}